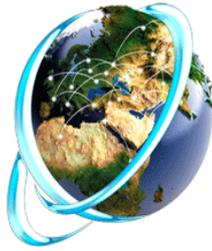

# A Novel Anonymous Cloud Architecture Design; Providing Secure Online Services and Electronic Payments

Alidoost Nia, Mehran; Master Software Student, Unversity of Guilan

Ghorbani, Aida; Master IT Student, Unversity of Guilan

Ebrahimi Atani, Reza; Asistant Professor, Department of Computer Engineering, Guilan University

**Abstract**

Anonymous cloud architecture provides secure environment for business and also e-commerce approaches. By using this type of architecture, we can propose anonymous online applications. Customers who need secure and reliable online services should pay for provided services. A big problem is electronic payment that is needed for billing customers. But customer identity should be remained anonymous during and also after payment procedure. In this paper we propose a novel and modified anonymous architecture that ensures customers that hide their identity from others. This architecture is used from common network protocols and we eliminate Tor anonymous service from architecture design space because of independency. The here is introduced scalability parameter in anonymous cloud architecture design space. After all we compare proposed architecture with other popular cloud architectures in this range and we obtain its advantages according to efficiency, security and anonymity.

## 1. Introduction

Today, online services and electronic payments are more common in online transactions. When cyber space replete with a lot of malicious treats, we should propose a new approach for safe network structures. But in this era with the growth of Internet and also private Intranets, It's more important to focus on secure interactions through the online payments.

Cloud computing is widely common in framework of organization structures. The main advantage of cloud computing is to reduce the cost of services and also enhance performance of the entire system overlay. Most of IT managers in government side, believe that performing cloud features in IT infrastructures can help to increase cooperation and collaboration among specific roles in system backbone. But they face with more challenges in implementation stage. One of the most important issues in cloud architecture design is the privacy and security features [19, 20].

Online service is a common and popular approach for attracting wide online users. Software as a Service is one of the best methods in cloud computing that can be implemented by cooperation of various services and provides real-time services through the network [2, 3]. The main idea is to provide secure and also anonymous online service among cloud computing infrastructure in specific organizations [6].

Security can be improved by many parameters like access control, anonymity, cryptography protocols and etc. however this is a trade-off between security enhancement level and system performance. Security consideration can be imposed on entire system working and leads to heavy burden on system processes. We need secure infrastructure; but the question is, how many parameters we need to perform? [12,15].

In a brief explanation we should attend to its applications. Electronic payment need to be executed in a secure way. In this specific application we want to focus on anonymous payment through insecure networks. When anonymity is applied, one of the most significant values in system will be improved. This value is called "trust". Trust feature can increase system reliability and also its reputation. In this paper we want to propose anonymous online services and payment structure via cloud architecture design. Secure architecture design plus anonymity guarantee lead to trusted and secure cloud infrastructure. In the next section, anonymous architecture design will be proposed and then it should be analyzed among security features. Then in section 4 we show its application in national and also private Intranets. Finally we discuss about its advantages and we compare it with other architectures.

## 2. Literature Review

Anonymization process is so important in security and reliable data exchange. But every work that leads to higher confidentiality surely increases data overload. This should be a motivation for moving toward lightweight cryptography algorithms [13].

Tor project is one of the perfect approaches for anonymous plans. This framework works according to onion router systems [7]. As shown in Figure1 messages should be encrypted by at least three layers. RSA, AES and also Deffi-helman key exchange technique are applied as cryptography mechanisms. Every node has its own public key and key exchange begins with RSA under public key. Then session key is constructed and after this, data transfer should be started by AES encryption. Layered encryption seems very good approach but we can't use Tor system for general applications with high QoS demand because of its payload and overhead that are imposed to processes [1].

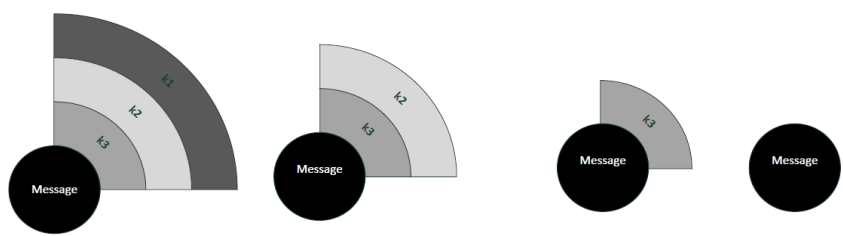

**Figure 1. layered encryption in onion routers**

SHALON is the more general way for anonymity based on open standards such as HTTP or SSL [18]. As mentioned. Tor encrypts both the link and the circuit layers on data transfer, but SHALON uses encryption only in tunnel layer that is cover application layer in network structure. We need a combination between Tor onion routers and common protocol support systems like SHALON. In this paper we are trying to focus on lightweight Anonymization that retains both customer and cloud location identity for secure applications like electronic payment and online services.

After Anonymization it's needed to know about proposed architecture management around network. Parts of the system never remember customer history. To ensure customer about this, we need to a reliable architecture. In former cloud architecture that is proposed in [16], manager was turned into a big problem because of its knowledge about customer. This was a dangerous problem versus anonymous structure. If an invader could access to manager data, all customer privacy was changed. As a solution for referred issue, we add a new structure to this architecture. This structure is derived from network management idea for using agents instead of manager [4, 5]. Agent plays roll as some duties of manager over this system that relates to customer privacy. This process will be killed after task accomplishment according to structure of Linux operating system by using process ID [21].

## 3. Architecture Design Space

Cloud computing is referred to cooperative services that have their own specific duties. previous architecture is proposed in [16]; but there is steel some problem relates to design space.in this section we introduce main concepts of anonymous architecture and then we will discuss about problems in implementation step. Figure 2 is shown a new version of architecture by new features in anonymous design space. All of these attempts lead to reform cloud architecture for Intranet and private usages. Concealing versus anonymity is an issue that can be handled by our new architecture.

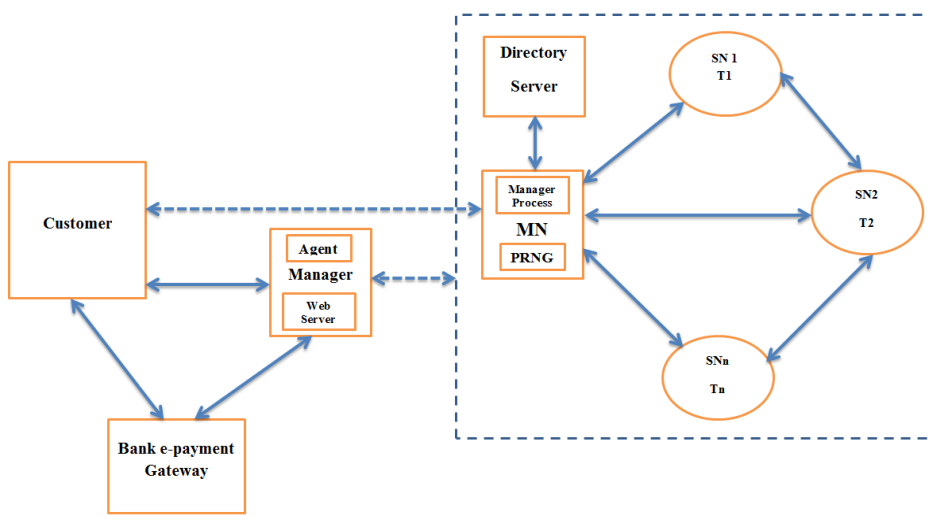

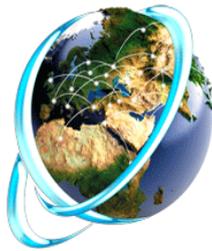

Figure 2. Agent-based & Tor independent anonymous cloud architecture

### 3.1. Roles of cloud architecture

*Customers* (C) who want to use cloud resources and pay for requests. This architecture has a guarantee that ensures us with two different policies. First is the anonymity of customer identity and second ensures about payments and billing process completely.

*Manager* (M) is responsible for receiving requests from customers and computes final cost of use by sending billing process to customers. Another duty of manager is to construct anonymous circuit for satisfying anonymity procedure. Our new approach is eliminated dependency of Tor system. Alternatively anonymous process for this architecture is indicated in the next section. This elimination seems very necessary because of Intranet and private applications. But one of the main advantages of Manager is agent. Agent process for every customer runs and after completing payment step, will be destroyed. Then no information about customer transaction is remained except some details related to its payment in bank gateway.

*Slave nodes* (SN) are capable to do customer sequence of jobs. In this approach, each SN must run anonymous routine plus its specific duties.

*Master node* (MN) is responsible for managing SNs after establishing anonymous connection through local Intranet and also after access control process and receiving authentication by manager. It can has unknown relationship with customer and do his jobs. Another duty is pseudonymity of SNs for running anonymity service. Also MN is responsible for running manager process for constructing circuits and provides agent requirements.

Directory server (DS) is responsible for providing anonymous structure for customer usages. DS is worked like Tor directory servers [11] but has a structural difference. In this architecture, DS in cooperation with MN generates directly circle of nodes. This sequence delivers to manager and anonymous connection will be established between customer and MN.

### 3.2. Anonymous Architecture

Before this, at the higher level and Internet design space we used Tor project application for providing anonymous system. But we propose this architecture for secure and private infrastructures like National Intranet of Iran or other private Intranets. Due to this motivation it's needed to independent anonymous architecture. It can be reduced complexity of design in comparison with Tor project because we use more trusted systems in local cloud architecture.

The only connection to out of local network is the manager. In this architecture is used public key cryptography for data exchange between SNs and is eliminated key agreement protocol like Diffi-helman because of its imposed burden. Customers after performing authentication mechanism and access control process and receiving tokens can start their adventure through secure cloud architecture, anonymously [14].

### 3.3. Reforms

Significant issue in anonymous plan was the manager knowledge about cloud details. If an invader attacks to the manager system and access to private data, all system structure details will be missed. But in this paper we propose a relationship between manager and cloud architecture, anonymously. There is steel other vulnerable probabilities like attacking to DS. But DS is used under cloud system and generally is anonymous.

For security enhancement about DS, we perform Pseudo Random Number Generator (PRNG) that is responsible for changing SN IDs and updating directory server list. This action will enhance security of entire system [9].

Another reform is using DS. This action leads to having cost-efficient system. Before this, using Tor system was decreased overall process speed of cloud architecture and with elimination of Tor we receive customer content at the higher level. Quality of service is one of the important parameters in system design that can be satisfied by appropriate performance.

Using common HTTP or SSL protocols for establishing connection through anonymous architecture, can lead to lower delay in packet transfer time. Also we combine this method with layered onion router systems. In addition, this architecture has both the high confidentiality of Tor and high performance of Shalon [17].

The main reform in this design is the reduction on manager knowledge about customers. It ensures the customer that its private data will be eliminated after its payment and only some metadata for proving customer identity will be remained through usages of services. All steps of this process are related to network management idea called SNMP [21]. According to SNMP that provides a protocol for communication between agent in manager side and manager process in MN side. After receiving customer request, agent sends requests to manager process and after granting permission, agent does customer jobs with sending a sequence of requests to manager process. When every task completes in agent, a notification will be sent to manager process. These processes will be killed in both MN and manager side, when jobs are finished.

## 4. Applications

The motivation for moving toward this type of cloud architecture is to have online services that work anonymous. The here, one of the most important application of this architecture is competition among companies. For example if a company wants to sell large portion of its stack, it will be more important to do this in silence. In other application, assume that a company wants to transfer large amount of its balance for business proposes in a secure way. By using this architecture, it can be guaranteed for maintaining private documents and sensitive actions of the company from rival companies.

Anonymous cloud also could be applied for general purposes like online storages, email and data transfer and online services such as e-payment, e-commerce and online software. Cloud network is one of popular services that is growth very fast in cloud infrastructure and emerged as NaaS. We can use this type of service through anonymous cloud. For example we can provide anonymous access to the internet in a local network for specific reasons. The main scenario for figure 2 is a customer that needs to use some online services in an unknown way. At the first step customer should go to webserver that is provided by manager. After choosing favorite services, customer should pay for services. Manager counts bill and redirects customer to bank gateway for e-payment step. Then bank will send a notification about customer payment details and agent process begins to prepare anonymous connection between cloud and customer. Finally, after customer requests are done, agent and manager process will be killed and all connection will be destroyed.

## 5. Security Analysis

Benefits of internal Intranet consist of service availability, strong authentication, nonrepudiation, detection, strong logging system and etc. With these benefits, organizations which use internal intranet could have safe communication with high quality. With combining Intranet and other secure systems we could have safer relation and communications. Protection of intranet and its data helps to preserve confidentiality and integrity, service availability and it has become an important and major challenge for organizations. In these conditions, for more security of especially financial transactions data and processes, secure systems such as anonymity systems could be the best choice for implementing more safety in communications. One of the best methods is to use anonymous cloud.

In this section we would analysis our system which is based on anonymous cloud system. As regards, anonymous cloud uses public key cryptography method based on anonymous authentication for billing and accounting role of the manager (M), M could has access some metadata such as data and service ownership. Without using secure encryption method which being used in Tor (main anonymous cloud ), attackers could have access to important data with link ability or making relations between services and persons whom are using the special service and information which are being sent from source or destination of the communication.

Anonymous cloud architecture shares only metadata with manager of system for billing and accounting process which is encrypted with Public Key of manager. So the manager could access to metadata and unwrap it from the main data. Tor system nodes are being located outside of local network and could be out of service or filtered, and these nodes could be harmful for our anonymous cloud. Hence we decided to use a new architecture without Tor nodes but have the same or better anonymity in communications and relations with an internal directory server which has two-way relation with M and MN. With this manner, system could be more secured than main anonymous cloud method.

Customer (C) sends request of using resources and services of anonymous cloud system and trying to communicate which is M. In this position, C asks token (T) for transactions, S (type of service and service number) from M and encrypt this data with $K_M$ (the public key of manager), so only manager could access to the data which is being asked. For constructing anonymous circuit, M tries to communicate with the internal DS and asks about the list of anonymity nodes, SNs and MN to build a secure circuit of anonymity.

Our new architecture is internal, so important information like location of SNs and MN or information which is traveled between SNs and MN was held in a proper security form, which couldn't be assured with anonymous manner. Internal DS sends list of nodes which is provided to serve those services which C asked, to the manager and M sends information which C needs to communicate with the anonymous cloud. There are no relations between C and DS, this point is advantage not disadvantage.

If important information like location of MN would be shared with C, It will face with some kind of information leakage because we decided not to have a relation between DS and C. Relation of DS and M would be secured if we use a trusted CA server and generate one way trusted certificate to have trust between DS and M. Using internal manager is one of the other benefits of our architecture. In our design, M in the organization is internal and its security is being assured by internal/external firewalls and physical methods.

M runs the process of servicing and sends the process ID to C for raising trust for C. Then the unique process communicates with MN to serve the process. These requests consist of token number, service request number and etc., which are being encrypted with $K_{MN}$ (the public key of MN). Only MN could access to the request. MN Process breaks the service into proper sub-services and sends them to SNs which are participating in the circuit of anonymity. To have a better anonymous communication, MN would use PRNG manner for SNs. After the process of servicing and serving, anonymously MN would send service and a message to the M which contains of servicing process is done and M could start billing process and sends bill to customer. If C wants to use another service, it should send another request of creating new session to the system and M would generate another process and unique ID for it. In this approach we don't use TOR, so to have better anonymous system we use PRNG manner in MN. PRNG makes pseudonym names for each SN in the circuit. When C receives the service which had asked and bill from M, should send acknowledgement of receiving then should pay the bill. The process after receiving the payment from C would be killed.

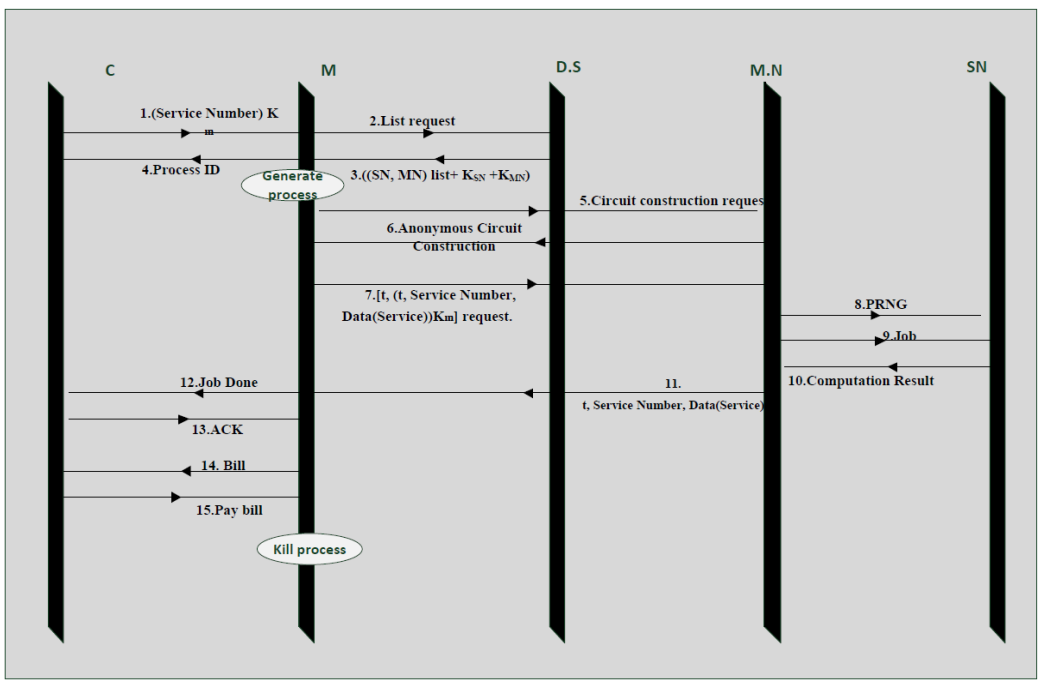

Figure 3. Secure communication steps

## 6. Comparison and Advantages

Platform independent service is the main motivation for moving toward cloud architectures. Online services like Google cloud and its online applications have high level of security. In this section we compare security and efficiency parameters in our novel design with other secure and reliable cloud architectures. One of important parameters that improved in comparison with other architectures is scalability. It means that we can use this type of architecture for small and also large scale of cloud networks. Assume that we have a small organization with local network that needs secure communication via cloud architecture for services in its own network. Our proposed architecture satisfies all requirements of organization by security features with minimum cost of usage.

As shown in Table1, we do brief comparison between our new architecture and current secure systems. Our novel approach decreases cost of overall system implementation. Also it offers high level and also scalable security model with emphasis on anonymous services. These specifications lead to more security with high efficiency. Also for more security we use PRNG system pseudonym assignment. It can improve anonymous communication and also hide server location for attack prevention plans.

As mentioned our manner would be implemented locally so it has high security than old technique. In this approach a unique process would be generated in manager system to do the job of computing which had been asked from system by customer. This process will be killed after serving the customer request of task so the information of process would be destroyed and no one could reach to them. Hence the anonymity of our approach is higher than old system.

Our architecture uses from common network protocols like http or SSL. This is an advantage that makes it compatible with network and especially internet applications. Tor project only uses its own servers and also depends on specific platform. Unlike Tor, Google cloud is provided via internet and significant point in Google design is high efficiency. As shown in Figure4, Google cloud uses multitenant technique for anonymity of storage services. This action helps the maintaining the security of data. In this technique data that should be stored in data centers, spreads into several portions and according to specific storage algorithm, data will be stored in different position. In overall Google anonymity is not high because of its applications that are not anonymous.

Table 1. The Comparison between our architecture and other current cloud systems

| Architecture | Level of anonymity | Efficiency | Level of security | scalability | Distributed environment | Anonymity technique | Anonymity applications |
|---|---|---|---|---|---|---|---|
| **Anonymous cloud with Tor [16]** | High | Middle | Middle | Low (only in small scales) | Is not used | Tor | e-payment online services |
| **Our architecture** | High | High (http or SSL) | Middle | High (both small and large scales) | Is not used | Layered Encryption | e-payment online services |
| **Google cloud [10]** | Low | High (via internet) | High | Low (only in large scales) | used | Multitenant | Storage |

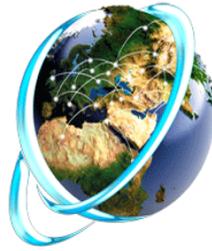

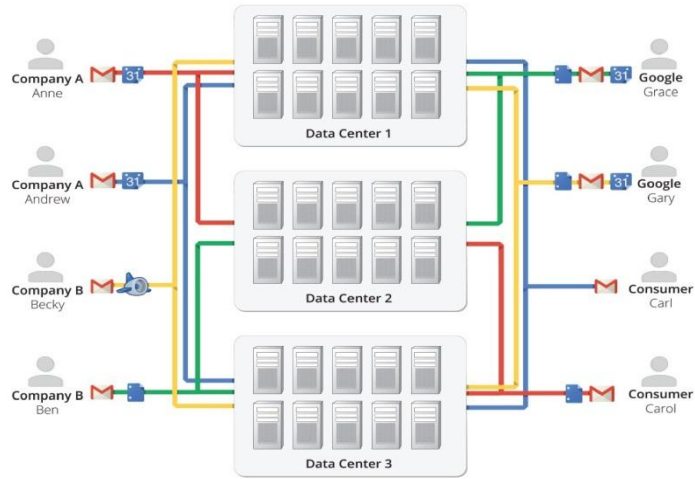

**Figure 4. Google multitenant storage system [10]**

## 7. Conclusion

Anonymous online services via cloud architecture are needed for online users who want to use from secure structure. Online services become popular by developing online applications. These applications have a large numbers of advantages like platform independency, cost of service, number of users and etc. But online users need high security level for their applications. Anonymous architectures improve level of reliability and also security of the entire system. So we are moving toward designing new and more secure architectures according to user requirements.

We propose a novel architecture that provides agent-based system process. This architecture has new and significant features like high scalability for both small and large scale organizations, high efficiency, Tor independent architecture, common protocol usages and appropriate for e-commerce applications. We eliminate key role of Manager in popular architectures and use from agents for message transportation. Customers can trust the entire system architecture and as a result we have a more reliable architecture. It works both with internet and without internet via local networks. Unlike Tor it will be large scale with large number of nodes and will be small scale according to small organizations. In this architecture we use Slave Nodes bidirectional. First providing anonymous services and second doing its duty along cloud system like providing common used services.

For future works we can use SSH-based layered encryption systems in cloud architecture. SSH is a strong approach for transporting data from one point to another. A new approach is to transfer data packages from one server to another (like onion routers) by changing IP number in servers. In fact we change IP and SSH port for transporting data packages [8].

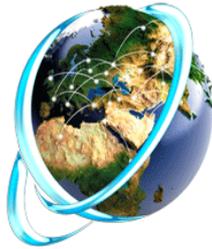